\documentstyle[preprint,aps,epsfig]{revtex}

\begin{document}

\titlepage
\draft

\preprint{
\vbox{
\hbox{ADP-98-79/T346}
}}

\title{Pion Exchange and the H1 Forward Spectrometer Data}
\author{Anthony W. Thomas 
and Csaba Boros} 

\address{Department of Physics and Mathematical Physics and \\
Special Research Centre for the Subatomic Structure of Matter \\
University of Adelaide, SA 5005, Australia}

\maketitle

\begin{abstract}
We point out that the $\Delta\pi$ component of the nucleon 
wave function is vital to the interpretation of the recent H1 data for 
leading baryon production. While the $n/p$ ratio is equal to two 
with the $N\pi$ component alone, the inclusion of the $\Delta\pi$ component 
brings this ratio very near to unity, as observed in the experiment. 
This result demonstrates that pion exchange can not only account for 
leading neutron  but also for  
a large fraction of the leading proton production.
\end{abstract} 

\newpage

\section{Introduction}
\label{intro}

Over the past few years data collected at DESY have vastly increased our
store of knowledge concerning nucleon structure functions. One
particular class of events, discovered by the ZEUS 
\cite{ZEUSGAP} and the H1 \cite{H1GAP} 
collaborations, has caused enormous interest. These are the ``rapidity
gap events'', which amount to some 10\% of the total deep inelastic
cross section. Events in this class are characterised by a large,
particle free gap in rapidity between the region of phase space occupied
by the debris of the target proton and the jet associated with the
interaction current. 

While these events certainly involve the Pomeron and have provided
important new information concerning its properties, it has also been
realised for some time that the pion cloud of the nucleon, required by
non-perturbative QCD because of dynamical symmetry breaking, may
play a role \cite{BOR95}.  
Although the rapidity gaps are much smaller in pion exchange   
than in Pomeron exchange \cite{INGEL},  
both are characterised by the production of   
fast baryons in the forward region.    
The pion cloud was first discussed in the context of deep
inelastic scattering by Feynman \cite{Fey} and Sullivan \cite{Sul}. It
was later realised that, as well as leading to an excess of non-strange
over strange sea quarks, the pion cloud would yield a significant excess
of $\bar{d}$ over $\bar{u}$ quarks in the proton \cite{Tho83}. 

This mechanism for violating the Gottfried sum rule, while preserving
isospin, has been extensively studied theoretically \cite{PiGott}
since the New Muon Collaboration discovered that 
the Gottfried sum rule was violated
\cite{NMC} -- for recent reviews see \cite{SpeT,LonT,Kum}. Later experiments
by NA51 (at CERN) \cite{NA51}, E866 (at Fermilab) \cite{E866} and most 
recently HERMES (at DESY) \cite{HER} have given us quite detailed
information on the shape of $\bar{d}(x)/\bar{u}(x)$ and it is clear
that the pion cloud plays an important role in understanding this data
\cite{MT98}. From the phenomenological point of view, once one can
establish the role of pions in this type of diffractive event  
one can 
use such data to study the pion structure function at small $x$
\cite{SpeNNN} -- something that is difficult to obtain any other way
\cite{LT}.

In order to specifically study the role of pions in the rapidity gap
events, the H1 detector was upgraded by the addition of a forward proton
spectrometer (FPS) and a forward neutron spectrometer (FNS). Both were
specifically designed to detect forward going hadrons with $p_T$ up to
200 MeV/c (recall that the beam momentum is 800 GeV/c!). The expectation
of the collaboration was that if pion exchange alone were 
responsible for leading baryon production, the ratio of $n$ to $p$ production
would be in the ratio 2:1 -- coming from the square of the isospin
Clebsch-Gordon coefficients for $p \rightarrow n \pi^+$ and 
$p \rightarrow p \pi^0$ ( $\frac{2}{3}$ and $\frac{1}{3}$,
respectively).

The results of the H1 measurements were released recently \cite{H1_98}. 
A major finding was that in the relevant region of phase space the
semi-inclusive proton production cross section was slightly larger than  
that for neutrons and that this ruled out pion exchange as the main
mechanism for leading protons. Our purpose is to point out that, while
the proton data does require other mechanisms as well, a large fraction
of the proton events can indeed be understood in terms of pion exchange.
We stress that the expectation of a 2:1 ratio is a little too naive and that
well established physics associated with the pion cloud of the nucleon
leads us to expect the experimental ratio to be closer to 1:1. 
While the role of the $\Delta$ in these processes was  discussed 
quantitatively by Szczurek et al.  
\cite{SpeNNN} the experimental analysis  totally 
omits any consideration of it. 
Our aim here has therefore been to 
specifically avoid the details of the
experimental acceptance, but concentrate 
on the essential physics of this
experiment. In this way we hope to focus attention 
on the need to reanalyse the data taking the effects of the 
$\Delta$ resonance into account.

\section{The pion cloud of the nucleon}
\label{sec:1}

A complete analysis of the H1 data requires a full Monte-Carlo calculation
including momentum acceptance cuts that can only be done by the
collaboration. Our purpose is to present some physics which has so far
been omitted from the analysis, which is nevertheless vital to
the interpretation of the data.

A full solution of QCD with dynamical symmetry breaking is still just a
dream for theorists. For the present we rely on a mixture of QCD
motivated models and phenomenology. Although there are now many
sophisticated chiral quark models of nucleon structure, it is often not
easy to appreciate the physics. The cloudy bag model (CBM) \cite{CBM}
is both physically transparent and produces a picture of the nucleon,
especially the probabilities for specific meson-baryon Fock states, that
is in remarkably close agreement with modern analyses of the meson
contribution to the spin and flavor structure of the nucleon -- see Ref.
\cite{SpeT}. While we use it to guide our discussion, we expect the
general features to be quite robust.

Under SU(6) symmetry the $N$ and $\Delta$ are degenerate and hence we
might expect to treat them on the same footing. In the CBM, even though
the $N-\Delta$ degeneracy is removed, this is still true. The
transitions $N \rightarrow N \pi$ and $N \rightarrow \Delta \pi$
do not change the orbital occupied by the active valence quark. As a
result the two processes have coupling constants that are large and
similar in magnitude. Under SU(6) symmetry the momentum dependence of 
the two vertex functions is identical -- in the CBM it is $3 j_1(kR)/kR$
which, for many practical purposes may be approximated by $e^{-k^2R^2}$,
with $R$ the bag radius. This seems phenomenologically reasonable
because the axial form factor of the nucleon and for the  $N \rightarrow
\Delta$ transition are very similar in shape \cite{Axial}. The
relatively large excitation energies and smaller coupling constants for
transitions to higher mass baryons suppress their contribution to
nucleon properties, so that in practice the major effects come from 
$N \pi$ and $\Delta \pi$ components of the wave function.

The dominant Fock components of the $p$, with their probabilities, are
therefore:
\begin{eqnarray}
\pi^+ n &:& \frac{2}{3} P_{N\pi} \hspace{0.5cm}; \hspace{0.5cm} \pi^0 p :
\frac{1}{3} P_{N\pi} \nonumber \\
\pi^- \Delta^{++} (&\rightarrow & \pi^+ p) \hspace{0.5cm} : 
\hspace{0.5cm} \frac{1}{2} P_{\Delta \pi}
\nonumber \\
\pi^0 \Delta^{+} (&\rightarrow & \pi^+ n / \pi^0 p : 
\frac{1}{3} /
\frac{2}{3}) \hspace{0.5cm} :  \hspace{0.5cm} 
\frac{1}{3} P_{\Delta \pi} \nonumber \\
\pi^+ \Delta^{0} (&\rightarrow & \pi^0 n / \pi^- p : 
\frac{2}{3} /
\frac{1}{3} ) \hspace{0.5cm} : \hspace{0.5cm} \frac{1}{6} P_{\Delta \pi}.
\end{eqnarray}
Based on experience with the CBM as well as the phenomenological analysis
of deep inelastic scattering data in the meson cloud model 
\cite{SpeT,MT98}, we expect the total probability of the $N \pi$ Fock component
($P_{N\pi}$) to be 18-20\%, while the $\Delta \pi$ probability
($P_{\Delta \pi}$) would be 6-12\%. 

We recall that the FPS and FNS limit $p_T$ to less than 200 MeV/c. Since
the vertex functions for $N \rightarrow N \pi$ and $N \rightarrow \Delta
\pi$ are approximately the same, as explained earlier, the distribution
of $N$'s and $\Delta$'s in $p_T$ will be essentially identical. Of
course, the $\Delta$ will decay well before reaching the forward
spectrometers. Most of the time this will produce a proton, and as the
typical transverse momentum in the decay of the $\Delta$ is also around
200 MeV/c these will mostly be detected by the FPS. Even in the case
where a $n$ is produced by the decay of the $\Delta$, the pion will pass
through the forward spectrometer and be vetoed by the FNS, thus counting
as a ``proton''.

The exact detection efficiencies are a matter for the experimental
group's Monte Carlo simulation. In order to estimate the effect of the
$\Delta \pi$ Fock component we make two assumptions: a) only the protons
produced by $\Delta$ decay will count as protons and anything else as a
neutron; b) any charged particle produced by delta decay will look like
a proton (since the FPS has no particle identification) and will be
counted as such. Under assumption (a) and using the coefficients given
in Eq. (1), the $n$ over $p$ ratio is:
\begin{equation}
R \left( \frac{n}{p} \right) = \frac{\frac{2}{3} P_{N\pi} + \frac{2}{9} 
P_{\Delta \pi}}{\frac{1}{3} P_{N\pi} + \frac{7}{9} P_{\Delta \pi}}.
\end{equation}
On the other hand, under assumption (b) we find:
\begin{equation}
R \left( \frac{n}{p} \right) = \frac{\frac{2}{3} P_{N\pi} + \frac{1}{9}
P_{\Delta \pi}}{\frac{1}{3} P_{N\pi} + \frac{8}{9} P_{\Delta \pi}}.
\end{equation}
In Table 1 we show the $n/p$ ratios for cases (a) and (b) for several
choices of the $\Delta \pi$ probability, ranging from 6 to 12\%. The
larger values are favoured by many analyses, but in fact the ratio is
not strongly dependent on it. It is always around unity and slightly
below unity at the prefered, upper 
end of the range. (We do not show the dependence on $P_{N \pi}$, because
the ratio is even less sensitive to that choice within the allowed
range.) It should be noted here also  
that these numbers serve as a first estimate. 
The decay of $\Delta$'s into nucleons will shift the energy distribution  
of these secondary particles to lower energy values  
decreasing these ratios somewhat at high energies. This effect 
is discussed in Ref. \cite{SpeNNN} and should be taken into account  
in the Monte Carlo simulations.  

It should be clear from this analysis that the $\Delta \pi$
component of the wave function of the nucleon is vital to the
interpretation of the H1 data, bringing the $n/p$ ratio very near to
unity, as observed in the experiment, rather than two.

\section{Discussion}
\label{disc}
 
We have seen that the $\Delta \pi$ component of the nucleon wavefunction
is vital to the analysis of the FPS and FNS data taken by the H1
collaboration. As we have emphasised this is not a unique example of its
importance. In analysing the violation of the Gottfried sum rule, and
more particularly the ratio of $\bar{d}/\bar{u}$, the $\Delta \pi$ and
$N \pi$ components tend to cancel each other and the detailed
description of the data requires a careful treatment of both components
\cite{MT98}. Within the CBM, the explicit presence of the $\Delta$ was
essential to the rapid convergence properties of the theory -- for
example, the fact that the bare and renormalized $NN\pi$ coupling
constants were typically within 10\% of each other \cite{CBM}.  
The left-right asymmetry data for inclusively produced 
pions, measured by the FNAL E704-Collaboration \cite{E704}  
using transversally 
polarized proton beams and unpolarized targets, also  
suggest the importance of the $\Delta \pi$ component in the 
nucleon wave function. The experimental observation that 
the asymmetry of $\pi^+$ and that of $\pi^-$ have different signs 
can be understood if one notes that  the spin of the 
baryon in the meson-baryon fluctuation determines the  
angular momentum  dependence of the wave function 
and that the lowest lying components relevant for the 
production of the leading $\pi^+$ and $\pi^-$ are the  
$N\pi$ and the $\Delta \pi$ components, respectively 
\cite{BOR98}.  We could  
cite many other examples but for the present we simply urge the collaboration
to include the $\Delta \pi$ component of the nucleon wavefunction in a
full Monte Carlo analysis of the data.

{\bf {ACKNOWLEDGMENTS}}
This work was supported in part by the Australian Research Council.

%

%
%

%
\begin{table}
\caption{Neutron to proton ratios under the scenarios described in the
text -- the $N \pi$ probability is chosen to be 18\%.}
\label{tab:1}       
\begin{tabular}{llll}
\hline\noalign{\smallskip}
Case & $P_{\Delta \pi}$ = 6\% & $P_{\Delta \pi}$ = 9\% & 
$P_{\Delta \pi}$ = 12\%  \\
\noalign{\smallskip}\hline\noalign{\smallskip}
(a) & 1.25 & 1.08 & 0.96 \\
(b) & 1.12 & 0.93 & 0.80\\
\noalign{\smallskip}\hline
\end{tabular}
\end{table}

\end{document}